\documentclass[aps,prd,eqsecnum,onecolumn,showpacs,nofootinbib,preprintnumbers,a4paper]{revtex4}
\usepackage{amsmath}
\usepackage{amssymb}
\usepackage{amsfonts}
\usepackage{latexsym}
\usepackage{graphicx}
\usepackage{dcolumn}
\usepackage{bm}
\usepackage{empheq}
\usepackage{color}
\usepackage{times}  
\newcommand{\eq}[1]{\begin{equation}#1\end{equation}}

\newcommand{\ea}[1]{\begin{equation}\begin{aligned}#1\end{aligned}\end{equation}}
\newcommand{\itm}[1]{\begin{itemize}#1\end{itemize}}
\newcommand{\lrp}[1]{\left( #1 \right)}  
\newcommand{\lrb}[1]{\left( #1 \right)}  
\newcommand{\lrsb}[1]{\left[ #1 \right]}  

\def\rd{\partial}
\def\vx{\bm{x}}
\def\vk{\bm{k}}

\definecolor{gxhighlight}{rgb}{1,1,0.4}
\renewcommand{\d}{\mathrm{d}}

\def\dop{\dot{\varphi}}

\def\doq{\dot{Q}}
\def\bea{\begin{eqnarray}}
\def\eea{\end{eqnarray}}
\def\be{\begin{equation}}
\def\ee{\end{equation}}

\begin{document}

\title{Fluctuations in a Ho\v{r}ava-Lifshitz Bouncing Cosmology}

\author{Xian Gao$^{1)}$, Yi Wang$^{2)}$, Wei Xue$^{2)}$ and Robert Brandenberger$^{2)}$}
\affiliation{1) Key Laboratory of Frontiers in Theoretical Physics, Chinese Academy of Sciences,
Beijing 100190, P.R. China}
\affiliation{2) Physics Department, McGill University, Montreal, QC, H3A 2T8, Canada}


\date{\today}



\begin{abstract}
Ho\v{r}ava-Lifshitz gravity is a potentially UV complete theory with
important implications for the very early universe. In particular, in
the presence of spatial curvature it is possible to obtain a
non-singular bouncing cosmology. The bounce is realized as a consequence
of higher order spatial curvature terms in the gravitational
action. Here, we extend the study of linear
cosmological perturbations in Ho\v{r}ava-Lifshitz gravity coupled to
matter in the case when spatial curvature is present. As in the case
without spatial curvature, we find that there is no extra dynamical
degree of freedom for scalar metric perturbations. We study the evolution
of fluctuations through the bounce and show that the solutions
remain non-singular throughout. If we start with quantum vacuum
fluctuations on sub-Hubble scales in the contracting phase, and
if the contracting phase is dominated by pressure-less matter,
then for $\lambda = 1$ and in the infrared limit the perturbations
at late times are scale invariant. Thus, Ho\v{r}ava-Lifshitz gravity
can provide a realization of the ``matter bounce'' scenario of structure
formation.
\end{abstract}

\maketitle \tableofcontents 

\section{Introduction}

Ho\v{r}ava has proposed a simple model of quantum gravity
~\cite{Horava0,Horava1} which is
conservative in the sense that it is based
on using the usual metric degrees of freedom in four space-time
dimensions, but is radical in the sense that it abandons general covariance
and local Lorentz invariance. Instead, the theory is based on
a scaling symmetry in which space and time scale differently.
Spatial diffeomorphism and space-independent time reparametrizations
remain as symmetries of the theory. Ho\v{r}ava-Lifshitz gravity,
as this theory is now called, has a free-field ultraviolet (UV) fixed
point. It is argued that there is also an infrared fixed point in
which the action reduces to that of General Relativity and in which
local Lorentz symmetry and space-time diffeomorphism invariance
emerge. There have been several general studies of
Ho\v{r}ava-Lifshitz gravity \cite{general} and a number of
studies of cosmological aspects of the theory \cite{cosmology}.

Since Ho\v{r}ava-Lifshitz gravity has the same number of fields
as General Relativity but has a reduced symmetry, we should expect
an extra physical mode \cite{Horava0}. This mode could be
ghost-like \cite{Saridakis2}, it could be strongly coupled
\cite{Charmousis,Blas}, or it could be simply well-behaved but
phenomenologically ruled out. In a previous paper \cite{us} (see also
\cite{also}) we showed that, in the absence of spatial curvature, the
extra mode is not propagating at all (this conclusion was later
confirmed in \cite{Maartens}). Our analysis showed that the
strong coupling instability discussed in \cite{Charmousis}
and the ghost-like evolution studied in \cite{Saridakis2} are
regulated by taking into account the expansion of space which
is inevitable in the presence of background matter \footnote{After
the work reported in this paper was completed, a paper
appeared \cite{Blas2} showing how the strong coupling
instability can be resolved by adding extra terms to the
original Ho\v{r}ava-Lifshitz action.}.

Earlier, it had been shown \cite{RHB} (see also \cite{Cai:2009in}) that in the presence of
spatial curvature it is possible to obtain a non-singular
bouncing cosmology. At the bounce point the expansion rate of
the universe vanishes and hence the question arises as to
whether linear cosmological fluctuations are well-behaved in
a bouncing Ho\v{r}ava-Lifshitz cosmology in the same way as
they are well-behaved in a spatially flat expanding cosmology.
In this paper we will firstly show that the presence of
spatial curvature does not change the conclusion that there
are no extra propagating degrees of freedom. Secondly, we
show that the fluctuations pass through the bounce smoothly
in spite of the fact that a term in the equations of motion
blows up.

As argued in \cite{RHB}, Ho\v{r}ava-Lifshitz gravity in the
presence of non-vanishing spatial curvature may yield a concrete
realization of the ``matter bounce'' (see \cite{Wands1,Fabio,Wands2}
for original works, \cite{recent} for more recent studies and
\cite{RHBrev} for a short review) scenario. In this scenario,
fluctuations which originate as quantum vacuum perturbations of
a matter scalar field on sub-Hubble scales in a matter-dominated
contracting phase will evolve into a scale-invariant spectrum of
curvature perturbations at later times in the expanding phase,
with a special shape and distinguished amplitude of the
three point function \cite{bounceNG}. However, in \cite{RHB}
the evolution of fluctuations was considered in the context
of the Einstein gravity equations, and without analyzing their
propagation through the actual bounce. The results of the
present paper show that the equations of General Relativity
indeed provide an excellent approximation to the actual
evolution for IR modes of interest to current cosmological
observations.

The outline of this paper is as follows. We first briefly review
Ho\v{r}ava-Lifshitz gravity. Next, we analyze the conditions which
must be satisfied in order to obtain a non-singular bouncing
cosmology. We find that in order to realize a non-singular bounce,
non-trivial spatial curvature (either a closed or an open universe)
is needed. We also specify the conditions on the matter content in
the contracting phase which must be satisfied in order to obtain
a bounce. In the next section we then extend the theory of linear
adiabatic cosmological perturbations \cite{us} to the case in which
spatial curvature is present. We show that there is no extra propagating
degree of freedom, as in the case studied in \cite{us} \footnote{Note that
we are considering the ``non-projectable" version of Ho\v{r}ava-Lifshitz
gravity.}. However, at
the bounce point some of the coefficients in the equations of motion blow
up. Thus, in section \ref{Pinb}, we study the evolution of
cosmological fluctuations through the bounce. We find that on IR scales
relevant for current cosmological observations the evolution of
fluctuations in the pre-bounce contracting phase is indistinguishable
from what happens in General Relativity. Then, we show that the
fluctuations evolve smoothly through the bounce. Finally, we show
that a scale-invariant spectrum of curvature perturbations emerges
in the case of a ``matter bounce'', i.e. a bouncing cosmology
in which the contracting phase is dominated by pressureless matter.
There are corrections of order $\lambda-1$, where $\lambda = 1$
is the IR fixed point at which the IR part of the action reduces to that
of General Relativity. Finally, we discuss our results and give some
conclusions.

\section{Brief Review of Ho\v{r}ava-Lifshitz theory}

In Ho\v{r}ava-Lifshitz gravity space and time are treated
differently. The space-time manifold has an extra
structure, namely a given foliation of space-time into constant
time hypersurfaces. Instead of full space-time diffeomorphism
invariance, the symmetry of the Ho\v{r}ava-Lifshitz theory is
foliation-preserving diffeomorphisms, which consists of
(time-dependent) spatial diffeomorphisms and space-independent
time reparametrizations. A key ingredient in the theory
is the anisotropic scaling symmetry
\be
t \rightarrow l^z t \ , \quad x^i \rightarrow l x^i \ .
\ee
In order to obtain a power-counting renormalizable theory of
gravity in four space-time dimensions we set the scaling dimension
$z=3$. In this case, the theory in the UV region should flow to a
free-field fixed point and is renormalizable
by power counting. Meanwhile, in the IR region the theory is expected
to flow to the General Relativity limit where $\lambda = 1$.

The basic variables are the spatial metric $g_{ij}$,
the shift vector $N^{i}$ and the lapse function $N$.
\be
ds^2 \, = \, -N^2 \d t^2 + g_{ij}(\d x^i +N^i \d t)(\d x^j +N^j \d t) \ .
\ee
The spatial metric and the shift vector are functions of space and time.
For the lapse function there are two choices: either $N$ depends only
on time (when the so-called ``projectability condition'' is
satisfied), or it is taken to depend on both space and time
(the general case). We will consider the general case
\footnote{As discussed in
\cite{Horava1,Miao} there might be problems in the general
case when attempting to quantize the theory.}.

The action of Horava-Lifshitz gravity contains a ``kinetic" part and
a ``potential" part,
\eq{
    S^g = S^g_{K} +S^g_{V} \, . }
The action contains
the terms consistent with the symmetries of the theory (in particular
spatial diffeomorphism invariance) and with the correct scaling
dimension. The kinetic part is given by
    \eq{{\label{g_k_action}}
    S^g_K =  \frac{2}{\kappa^2} \int dt d^3x \sqrt{g} N
   \left( K_{ij}K^{ij}-\lambda K^2 \right)   \ ,
    }
where
\eq{
K_{ij} = \frac{1}{2N} \lrp{\dot{g}_{ij}- \nabla_i N_{j} - \nabla_j N_{i} } \, ,
    }
is the extrinsic curvature and $K= g^{ij}K_{ij}$. In General Relativity,
general covariance requires $\lambda=1$. The coupling constant $\lambda$
is dynamical and thus runs as the energy scale changes.

We will take the potential part of the action
to be of the ``detailed-balance" form
    \ea{
    S^g_V & = \int dt d^3x \sqrt{g}  N  \lrsb{ -\frac{\kappa^2}{2 w^4}C_{ij}C^{ij}
     +\frac{\kappa^2\mu}{2 w^2}\epsilon^{ijk}R_{il} \nabla_j R^{l}_{k}
       -\frac{\kappa^2\mu^2}{8}R_{ij}R^{ij} +\frac{\kappa^2\mu^2}{8(1-3\lambda)}\left(\frac{1-4\lambda}{4}R^2+\Lambda R-3\Lambda ^2\right)
     }
     \,,
    }
where $C_{ij}$ is the Cotton tensor defined by
    \eq{
        C^{ij}= \frac{\epsilon^{ikl}}{\sqrt{g}} \nabla_k \lrp{
        R^{j}_{l}-\frac{1}{4}R\delta^{j}_{l} }
        \,.
    }
Here and in the above, tensors like $R$ are understood to be
constructed from the spatial metric, and $g$ is the determinant
of the spatial metric.
The ``detailed balance'' condition reduces the number of terms in the
potential. The most general potential is discussed in \cite{Silke}.
Choosing the simple form of the potential will simplify our equations
(which are already complicated enough) but will not change our
basic conclusions concerning the number of dynamical degrees of
freedom and concerning the non-singular behavior of the solutions
through the bounce.

We consider the simplest form of matter to be coupled to gravity,
namely a scalar matter field \footnote{As is well known and as is reviewed
at the beginning of Section V, a scalar field oscillating about the minimum of
its potential yields a matter-dominated equation of state provided that the
quadratic term in the expansion of the potential about the minimum does not
vanish.}.
The general structure of the action of scalar-field matter in
Ho\v{r}ava-Lifshitz theory contains two parts: a quadratic kinetic
term invariant under foliation-preserving diffeomorphisms and a
potential term:
\eq{{\label{scalar_action}}
    S^{\phi} = \int dt d^3x\, \sqrt{g}N
    \lrsb{ \frac{1}{2N^2} \lrp{ \dop - N^i \rd_i \varphi }^2
    + F(\varphi,\rd_i\varphi,g_{ij})
        }\,,
    }
where $F$ will contain higher order terms in spatial derivatives
consistent with the symmetries and with power-counting renormalizability.

The speed of light in Ho\v{r}ava-Lifshitz theory can be obtained
by comparing the action with that of General Relativity.
The Einstein-Hilbert action in $3+1$ dimnsions is written in ADM form
as
\eq{
S_{EH} = \frac{c^3}{16\pi \hbar G} \int c dt d^3 x \sqrt{g} N
\{\frac{1}{c^2}(K_{ij}K^{ij}-K^2)+R -2\frac{\Lambda_{GR}}{c^2}\}
}
The expressions for the gravitational constant and the speed of light
in Ho\v{r}ava-Lifshitz gravity can be derived by comparing
the coefficients in the action with those in General Relativity.
In the infrared limit one obtains
\eq{
c = \frac{\kappa^2 \mu }{4} \sqrt{\frac{\Lambda}{1-3\lambda}} \ ,\
}
which can be seen from the ratio of the coefficients of the kinetic term
and the $R$ term. In addition,
\eq{
  16\pi G = \frac{\kappa^4 \mu }{8} \sqrt{\frac{\Lambda}{1-3\lambda}} \ , \
}
and
\eq{
\Lambda_{GR} = \frac{3\kappa^4 \mu^2 \Lambda^2}{32(1-3\lambda)}
= \frac{3}{2}c^2 \Lambda \, .
}
Finally, it is easy to get the coefficent of the $R^2$ term:
\eq{
\kappa^2 \mu^2 = \frac{8(1-3\lambda)c^3}{16\pi G\Lambda} \, .
}

\section{Matter Bounce Background}

In this section we analyze the background cosmology of
Ho\v{r}ava-Lifshitz gravity and study under which conditions a
non-singular bounce will occur.

We take the background metric to be
\begin{align}
  ds^2 = - dt^2+ \bar{g}_{ij} dx^idx^j  ~,
\end{align}
with
    \eq{
        \bar{g}_{ij} = a^2 \bar{h}_{ij} = \frac{a^2}{\lrb{1+\frac{\bar{k}
        r^2}{4}}^2} \delta_{ij} \,,
        }
where  $r^2 \equiv \delta_{ij}x^ix^j$ and $\bar{k}$ is the spatial
curvature which takes the values $\bar{k} = -1,0,1$. As we will see,
in order to obtain a matter bounce in Ho\v{r}ava-Lifshitz gravity,
$\bar{k}\neq 0$ is needed. Note that we are using units in which
the spatial coordinates $x_i$ are dimensionless (with respect
to the usual dimensions - not the anisotropic scaling dimension)
but the scale factor carries dimension of length.

The background equations of motion take the form
    \ea{ \label{friedmann}
    \frac{6(3\lambda-1)}{\kappa^2}H^2 &=
\rho-\frac{3\kappa^2\mu^2}{8(3\lambda-1)}\left(\frac{\bar
    k}{a^2}-\Lambda\right)^2~,\\
\dot \rho+3H(1+w)\rho&=0~, }
where $\rho$ and $p$ are the energy and pressure densities,
respectively, and the equation of state parameter $w$ is $w\equiv p/\rho$.
All other background equations are
consistent with the above equations, for example
\begin{align}
  \frac{2(3\lambda-1)}{\kappa^2}\dot
  H=-\frac{(1+w)\rho}{2}+\frac{\kappa^2\mu^2}{4(3\lambda-1)}\left(\frac{\bar
    k}{a^2}-\Lambda\right)\frac{\bar k}{a^2}~.
\end{align}
{F}rom (\ref{friedmann}) it follows that a bounce can only occur if
${\bar{k}} \neq 0$ (since otherwise $H = 0$ cannot be obtained).

In this work, we consider scalar field matter. The background energy $\rho$ and
pressure $p$ for this matter take the form
\begin{equation}
  \rho = \frac{\dot\varphi_0^2}{2}+V~,\quad p = \frac{\dot\varphi_0^2}{2}-V~.
\end{equation}
Since the cosmological constant must be tuned to be very small today,
we will concentrate on the case when $\bar k/a^2 \gg \Lambda$
is always satisfied.

When the equation of state for the scalar field satisfies $w < 1/3$,
then in the contracting phase the higher order curvature term in
\eqref{friedmann} will eventually catch up with the energy density
$\rho$, resulting in a cosmological bounce, a time when $H = 0$
and ${\dot H} > 0$.

To take one step further, we would not like to have super-deflation or
super-inflation around the bounce. For this purpose, we need
$\dot H$ to change sign twice, once before and once after the bounce time
when $H=0$. This is achieved if
\begin{align}
\left(\frac{\bar k}{a^2}-\Lambda\right)^2 <
\frac{4}{3(1+w)}\left(\frac{\bar k}{a^2}-\Lambda\right)\frac{\bar
k}{a^2}
\end{align}
which for negligible cosmological constant is realized is $w < 1/3$.
Otherwise, the cosmology will either begin with a phase of
super deflation leading to a bounce and then to deceleration, or
with accelerated contraction followed by a bounce and then super
inflation. Note that the ``matter bounce'' conditions
$\bar k/a^2 \gg \Lambda$ and $w=0$ yield a usual bounce without
super deflation or super inflation.

There exist three different phases in a matter bounce cosmology:
the first is the contracting phase during which the equation of
state is dominated by pressure-less matter. This phase ends when
the spatial curvature-induced higher derivative terms in the equations
of motion become important. When this occurs, the second phase -
the bouncing phase - begins during which the curvature-induced terms
will allow the universe to evolve from contraction to expansion in a
non-singular way. The last phase begins when the curvature-induced
higher derivative terms cease to be important as the universe grows
in size. At that point, the expanding phase that we observe today
begins. In order to be consistent with late time cosmology, there
needs to be entropy generation during or after the bounce such that
we get an expanding radiation phase. How to generate the
required amount of entropy is an issue we will not address here.

In the contracting and expanding phases, the scale factor can be
parameterized as a power law:
\eq{
a=a_B\eta ^{\frac{2}{1+3w}} \ ,
}
which yields
\eq{
\mathcal{H} = \frac{2}{(1 + 3 w)(\eta-\tilde{\eta_B})} \ ,
}
where ${\tilde{\eta_B}}$ is the time of the bounce. Note that
for a matter-dominated contracting phase the equation of state
parameter is $w=0$.

As in \cite{recent}, we model the evolution of the Hubble parameter
in the bouncing phase by linearly expanding in time about the bounce
point:
\eq{ \label{bounce}
a(\eta ) = \frac{a_B}{1-y(\eta-\tilde{\eta_B}) ^2}
}
which leads to
\eq{
\mathcal{H} = \frac{2y(\eta-\tilde{\eta_B}) }{1-y(\eta-\tilde{\eta_B}) ^2}
}

In the following sections we will study the evolution of linear
cosmological fluctuations in this background. We will assume that
the bounce occurs at a radius which is large in Planck units (this is
a natural assumption if the universe starts out cold and with a
length scale related to the initial temperature by dimensional analysis).
Later on in the text we will call this the ``large bounce radius assumption".


\section{Perturbations in the Presence of Curvature}

We will focus on scalar metric perturbations. In General Relativity,
these fluctuations can be described in terms
of four scalar functions of space and time $\phi, \psi, B$ and $E$
(see e.g. \cite{MFB} for a comprehensive review of the theory of
cosmological perturbations and \cite{RHBrev1} for a shorter overview):
\be
ds^2 \, = \, - (1 + 2 \phi) dt^2 + 2 \nabla_i B a(t)^2 dt dx^i
+ a(t)^2 \bigl[ (1 + 2 \psi) \delta_{ij} + 2 \nabla_i \nabla_j E \bigr]
dx^i dx^j \, .
\ee
There are two scalar gauge degrees of freedom which allow us to
elimate two of these four functions. For example, in longitudinal
gauge one chooses to set $E = B = 0$. However, in Ho\v{r}ava-Lifshitz
gravity one loses one of the gauge degrees of freedom, namely the
one corresponding to space-dependent time reparametrizations. Thus,
one can only eliminate one of the scalar degrees of freedom and one
should expect an extra propagating mode.

We will follow \cite{us} and use the remaining gauge freedom in the
scalar sector to eliminate the function $E$. Thus, we write the
perturbed spatial metric in the form
 \eq{{\label{metric_pert_form}}
    g_{ij} \equiv (1-2\psi) \bar{g}_{ij} = a^2\frac{ \lrp{1-2\psi} \delta_{ij} }{ \lrp{1+ \frac{\bar{k}}{4} r^2 }^2
    } \,.
}
Due to the conformal properties of the Cotton tensor, for the
perturbed metric (\ref{metric_pert_form}) $C_{ij} = 0$ and
$\epsilon^{ijk}R_{il} \nabla_j R^{l}_{k} =  0$.

In addition to the fluctuation in the spatial metric, there are
perturbations of the shift vector, the lapse function, and
the matter scalar field:
    \ea{
        N &= 1 + \phi(t,\vx) \,,\\
        N_i &= \nabla_i B(t,\vx) \,,\\
        \varphi &= \varphi_0 + Q(t,\vx) \,.
    }
Note that we are not enforcing the ``projectability condition''.
If we had enforced this condition, then $\phi$ would be constrained
to be a function of time only, and we could use the residual gauge
freedom of space-independent time reparametrizations to set
$\phi = 0$.

\subsection{Solving the Constraints}

The equations of motion for $N$ and $N_i$ are:
\ea{{\label{constraint}}
    0 &= -\frac{2}{\kappa^2} \lrp{ K_{ij}K^{ij} - \lambda K^2 } -
     \frac{\kappa^2}{2 w^4}C_{ij}C^{ij}
     +\frac{\kappa^2\mu}{2 w^2}\epsilon^{ijk}R_{il} \nabla_j R^{l}_{k}
       -\frac{\kappa^2\mu^2}{8}R_{ij}R^{ij} \\
       &\qquad\qquad + \frac{\kappa^2\mu^2}{8(1-3\lambda)}\left(\frac{1-4\lambda}{4}R^2+\Lambda R-3\Lambda
       ^2\right) -\frac{1}{2N^2} \lrp{ \dop -N^i\rd_i \varphi }^2 + F
       \,,\\
    0 &= \frac{4}{\kappa^2} \nabla_j\lrp{ K^j_i -\lambda K \delta^j_i } -
    \frac{1}{N} \lrp{ \dop - N^i \rd_i \varphi } \rd_i \varphi \,.
}
{F}or the background metric (\ref{metric_pert_form}), $C_{ij}=0$ and
$\epsilon^{ijk}R_{il} \nabla_j R^{l}_{k} =0 $.

At first-order, the energy constraint gives
    \ea{
        0 &=  \frac{4 (1-3 \lambda )H  }{\kappa ^2}\Delta B + \phi  \left(\frac{12 H^2 (1-3 \lambda )}{\kappa ^2}+\dop_0^2\right)  \\
        &\qquad +\frac{\kappa ^4 \left(\bar{k}-a^2 \Lambda \right) \mu ^2 \left(a^2 \Delta \psi +3 \bar{k} \psi \right)-2 a^4 (-1+3 \lambda ) \left(12 H (-1+3 \lambda ) \dot{\psi }+\kappa ^2 \left(\dot{Q} \dop_0+Q V'\right)\right)}{2 a^4 \kappa ^2 (-1+3 \lambda )} \,,
    }
while the momentum constraint yields
    \eq{
        0 = \frac{4}{\kappa^2}  \lrsb{ (-1+3 \lambda ) \left(H \phi +\dot{\psi }\right) -  \lrp{ \frac{2\bar{k}}{a^2} B + (1- \lambda)  \Delta
        B
        } } -   \dop_0 \, Q \,.
    }
In the above $\Delta $ is the Laplacian constructed using the background
spatial metric $\bar{g}_{ij}$.

As was done in the spatially flat model in \cite{us}, we
can combine the above two constraint equations and solve
(after choosing proper boundary conditions) for two of the
four fluctuation fields. We obtain
    \ea{{\label{constraint_sol}}
        \phi & = \frac{1}{2 a^4 (-1+3 \lambda ) \left(8 H^2 \left(3 \bar{k}+a^2 \Delta \right) (-1+3 \lambda )+\kappa ^2 \left(-2 \bar{k}+a^2 \Delta  (-1+\lambda )\right) \dop_0^2\right)} \\
        &\qquad \times \left\{ -16 a^4 H \left(3 \bar{k}+a^2 \Delta \right) (1-3 \lambda )^2 \dot{\psi } \right. \\
        &\qquad\qquad  +\kappa ^2 \left[ 2 a^4 (-1+3 \lambda ) \left(a^2 H Q \Delta  (-1+3 \lambda )+\left(-2 \bar{k}+a^2 \Delta  (-1+\lambda )\right) \dot{Q}\right) \dop_0 \right. \\
        &\qquad\qquad \left.\left. +\left(2 \bar{k}-a^2 \Delta  (-1+\lambda )\right) \left(\kappa ^2 \left(\bar{k}-a^2 \Lambda \right) \mu ^2 \left(a^2 \Delta\psi +3 \bar{k} \psi \right)+2 a^4 Q (1-3 \lambda ) V'\right)\right] \right\}  \,,\\
        B & = -\frac{\kappa^2 }{4 a^2 \left(8 H^2 \left(3 \bar{k}+a^2 \Delta \right) (-1+3 \lambda )+\kappa ^2 \left(-2 \bar{k}+a^2 \Delta  (-1+\lambda )\right) \dop_0^2\right)} \\
        &\qquad \times \left\{  4 a^4 H (-1+3 \lambda ) \left(3 H Q+\dot{Q}\right) \dop_0+4 a^4 (-1+3 \lambda ) \dot{\psi } \dop_0^2-a^4 Q \kappa ^2 \dop_0^3 \right. \\
        &\qquad\qquad \left. + 2 H \left[ \kappa ^2 \left(-\bar{k}+a^2 \Lambda \right) \mu ^2 \left(a^2 \Delta\psi +3 \bar{k} \psi \right)+2 a^4 Q (-1+3 \lambda ) V'\right]  \right\}  \,.
    }
The above solutions should be understood in momentum space where $\Delta
\equiv -k^2/a^2$. That is, we decompose the perturbations into
eigenfunctions $Q_{\vk}(\vx)$ of the background spatial Laplacian:
    \eq{
        \lrb{\Delta + \frac{k^2}{a^2}} Q_{\vk}(\vx) = 0\,,
    }
with eigenvalues
    \ea{
         \left\{ \begin{array}{cc}
                        k^2 \geq 0\,,& \quad \bar{k}=0 \\
                        k^2= \ell(\ell +2 )\,,& \quad \bar{k}= + 1 \\
                        k^2>1\,, & \quad \bar{k}=-1
                      \end{array}
          \right.
    }
Note that there is no singularity at the bounce point because ${\dot{\varphi_0}} \neq 0$
at the bounce time except for a measure zero set of initial conditions on the phase of
oscillation of $\varphi_0$.

\subsection{Quadratic Action}

Using (\ref{constraint_sol}), we get a quadratic action for the two
variables $\psi$ and $Q$:
\ea{{\label{S2_psi_Q}}
     S_2[\psi,Q] &= \int dt d^3x\, \sqrt{\bar{g}} \Big[ c_{\psi}\, \dot{\psi}^2 + c_Q\, \dot{Q}^2 + c_c\, \dot{\psi} \dot{Q} \\
     &\qquad\qquad\qquad\qquad  + f_{\psi}\, \dot{\psi}\psi + f_{Q}\, \doq Q + f_{c}\, \dot{\psi} Q + \tilde{f}_c\,\doq \psi    + \omega_{\psi} \psi^2 + \omega_Q\,
     Q^2 + \omega_c \, \psi Q
     \Big] \,,
}
where
    \eq{
        \bar{g} \equiv \det \bar{g}_{ij} = \frac{a^6}{\lrb{1+\frac{\bar{k}
        r^2}{4}}^6} \,,
        }
and the various ``coefficients"
(whose explicit expressions are given in Appendix \ref{appsec_coeff_S2_1})
should be understood in momentum space.

From Appendix \ref{appsec_coeff_S2_1} we notice that
\be
    c_{\varphi}\, \doq^2 + c_{\psi}\,
        \dot{\psi}^2 + c_{c}\, \doq \dot{\psi} \quad \propto
        \quad
        \lrp{ \dot{\psi} + \frac{H}{\dop_0} \doq }^2 \, ,
\ee
which  means that there is in fact only one dynamical degree of
freedom in our system. This degree of freedom is precisely the
Sasaki-Mukhanov \cite{SasMuk} combination of matter and metric perturbations,
defined as
    \eq{{\label{def_zeta}}
        -\zeta \equiv \psi + \frac{H}{\dop_0} Q \,,
    }
which is the gauge-invariant curvature perturbation on
uniform-density hypersurfaces
\footnote{Note that there is a singularity in the defining equation for $\zeta$
at times when ${\dot{\varphi}}_0 = 0$. This singularity is due to the fact that
at these times the uniform density hypersurface becomes degenerate and
hence $\zeta$ ceases to be a good variable to describe the fluctuations. This
problem also arises during reheating in inflationary cosmology and in that
context was studied in detail in \cite{FF,Zhang} with the conclusion that
$\zeta$ continues through this singularity without any problem.}.
From (\ref{def_zeta}), we can express
$Q$ in terms of $\psi$ and $\zeta$,
    \ea{
        Q &= -\frac{\dop_0}{H}(\zeta+\psi) \,,\\
        \doq &= -\lrp{ \frac{\ddot{\varphi}_0 H - \dop_0 \dot{H}}{H^2}
        }(\zeta+\psi) - \frac{\dop_0}{H}\lrp{\dot{\zeta} +
        \dot{\psi}} \,,\qquad \textrm{etc.}
    }
After plugging the above relations into (\ref{S2_psi_Q}), using
the background equations of motion and performing many integrations
by part, we get a new action for the two variables $\zeta$ and $\psi$:
    \ea{{\label{S2_zeta_psi}}
        S_2[\zeta,\psi] &= \int dt d^3x\,\sqrt{ \bar{g}} \lrb{ c_{\zeta}\, \dot{\zeta}^2 +  f_{\zeta} \, \dot{\zeta}\zeta  + \bar{f}_c \dot{\zeta}\psi  + \omega_{\zeta}\,\zeta^2
        + \bar{\omega}_c \zeta\psi + \bar{\omega}_{\psi} \psi^2
        } \\
    }
where the various coefficients can be found in Appendix
\ref{appsec_coeff_S2_2}.

{F}rom (\ref{S2_zeta_psi}), it follows that $\psi$ is not an independent dynamical
variable but rather a pure constraint, which can be
solved for explicitly in terms of $\zeta$
    \eq{{\label{psi_sol}}
        \psi = - \frac{ \bar{\omega}_c\, \zeta  + \bar{f}_c\, \dot{\zeta }}{2 \bar{\omega}_{\psi}}  \,.
    }
Note that the coefficient $\bar{\omega}$ does not vanish in our background. Hence, there
is no strong coupling instability related to the constrained field $\psi$.

Plugging (\ref{psi_sol}) into (\ref{S2_zeta_psi}), we get an
effective second-order action for a single variable $\zeta$
    \ea{
        S_2[\zeta] = \int dt d^3x \sqrt{ \bar{g} } \lrb{ \Gamma\, \dot{\zeta}^2 + f \,\dot{\zeta}\zeta
        + \omega_{HL} \, \zeta^2
        } \,,
        }
with
    \ea{
        \Gamma &\equiv c_{\zeta} - \frac{\bar{f}_c^2 }{4 \bar{\omega}_{\psi} } \,,\\
        f &\equiv  f_{\zeta} - \frac{\bar{f}_c\, \bar{\omega}_c }{2 \bar{\omega}_{\psi} }  \,,\\
        \omega_{HL} &\equiv  \omega_{\zeta} - \frac{\bar{\omega}_c^2}{4\bar{\omega}_{\psi}} \,.
    }
 The reader can verify that in the limit $\lambda = 1$ and for vanishing spatial
 curvature the coefficient $\Gamma$ reduces to that in the general relativistic
 theory of cosmological perturbations

After integrating by parts, we have
    \ea{
        S_2[\zeta] &\simeq \int dt d^3x \sqrt{ \bar{g} } \lrb{ \Gamma\, \dot{\zeta}^2  -\Omega\, \zeta^2
        } \,,
        }
where we made use of the definition
    \eq{
        \Omega \equiv - \lrsb{ \omega_{HL} -\frac{1}{2} \lrb{ \dot{f} +3H f }
        } \,.
    }

In order to write the action in canonical form, we introduce the new
variable
    \eq{{\label{udef}}
        u = z\,\zeta \,,
    }
with $z= a \sqrt{2 \Gamma}$. After changing to conformal time $\eta$
(which is defined by $dt = a d\eta$) we have
    \ea{
        S_2[u] &= \int d\eta d^3x\,\sqrt{\bar{h} }\, \frac{1}{2} \lrsb{ u'^2 + \lrb{
        \frac{z''}{z}
        - \frac{a^2 \Omega}{\Gamma} } u^2  } \,,
    }
where $\mathcal{H} \equiv a'/a$, $\bar{h} \equiv \det \bar{h}_{ij}$, $h_{ij}$ is
the background spatial metric without the factor $a^2$, and a prime indicates
the derivative with respect to conformal
time. Note that the above result should be understood in momentum space.

The classical equation of motion for the canonically normalized
variable $u$ is simply
    \eq{{\label{master_equation}}
         u_k'' + \omega^2(\eta,k)\,  u_k =0 \,,
    }
with
    \ea{{\label{omega}}
        \omega^2(\eta,k) &\equiv \frac{a^2 \Omega}{ \Gamma} -\frac{z''}{z} \\
        &=  \frac{a^2 \Omega}{ \Gamma} -\lrp{ \mathcal{H} + \frac{\Gamma'}{2\Gamma} }^2
          - \lrp{ \mathcal{H} + \frac{\Gamma'}{2\Gamma} }'  \, .
    }
As a consistency check, it can be verified that in the limit of General Relativity one
obtains the usual equation of motion found e.g. in \cite{MFB}.

\section{Evolution of Perturbations during the Bounce}
\label{Pinb}

In this section we will study the evolution of fluctuations from the time of their
generation early in the contracting phase until late times in the expanding
period. We will first review the evolution of fluctuations in a matter-dominated
phase of contraction in General Relativity. Then, we study the changes to the
evolution in the matter-dominated contracting phase which arise when the
dynamics is studied using the equations of Ho\v{r}ava-Lifshitz gravity.
The third step is to study the dynamics of the fluctuation modes in the bounce
phase, the transition period between matter-dominated contraction and
matter-dominated expansion. Finally, we need to match the solutions in the
bouncing phase to those in the post-bounce matter period.

\subsection{Einstein Gravity Analysis of the Matter-Dominated Contracting Phase}

Since we (for the sake of simplicity) model matter in terms of a scalar field,
we describe the background matter in terms of a scalar field condensate
$\varphi_0$.

In the limit $H^2 \ll m^2$ and making use of the WKB approximation, the
equation of motion for $\varphi_0$
\eq{
 \ddot{\varphi_0}+3H \dot \varphi_0 +m^2 \varphi=0 \ ,
}
can be solved, and the solution is
\eq{
 \varphi_0 \, \propto \, m^{-1/2}a^{-3/2}\exp (i \int m dt) \, ,
}
and thus we see that the energy density is propotional to
$a^{-3}$ and the time average of the pressure is approximately equal to zero.
Thus, the oscillating scalar field condensate indeed gives us a matter-dominated
contracting background cosmology.

We now turn to the description of the curvature fluctuations in the
matter-dominated contracting phase, first making use of the perturbation
equations from General Relativity. In this case, the equation of motion
for the canonical fluctuation variable $u$ defined in (\ref{udef}) reduces to
\be
u_k'' + \omega^2(\eta,k)\,  u_k \, = \, 0 \,,
\ee
with
\be \label{omegaeq}
\omega^2 \, = \, k^2 - \frac{z''}{z} \, ,
\ee
where $z \propto a$ as long as the equation of state of the background is unchanged. In
a matter-dominated phase, then on scales much larger than the Hubble radius (where the
first term on the right-hand side of (\ref{omegaeq}) can be neglected) we have
\be
\omega^2 \, = \, - \frac{2}{\eta^2} \, .
\ee

If the fluctuations originate as quantum vacuum perturbations on sub-Hubble scales
early in the contracting phase, then at Hubble radius crossing we have a vacuum
power spectrum $P_u$ for $u$, i.e.
\be
u_k(\eta_H(k)) \, = \, \frac{1}{\sqrt{2 k}} \, ,
\ee
where $\eta_H(k)$ is the conformal time when the scale $k$ exits the Hubble
radius, and hence
\be
P_{u}(\eta_H(k)) \, \sim \, k^2 \, .
\ee
To convert this vacuum spectrum into a scale-invariant one, we require a mechanism
which boosts the amplitude of long wavelength modes relative to those of short
wavelength ones.

In an expanding universe, the amplitude of the dominant mode of $u$ on
super-Hubble lengths grows as $z(\eta)$ and hence the curvature fluctuation
$\zeta$ is constant. In contrast, in a contracting phase the dominant mode of
$u$ grows as $\eta^{-1}$ (and hence $\zeta$ scales as $z^{-1}\eta^{-1}$). This
provides exactly the boost of long wavelength modes required to turn the initial
vacuum spectrum of fluctuations into a scale-invariant one, as can be seen
as follows:
\be \label{powerspec}
P_{\zeta}(k, \eta) \, = \, z^{-2}(\eta) P_u(k, \eta) \, = \,
z^{-2}(\eta) \bigl(  \frac{\eta_H(k)}{\eta} \bigr)^2 \, ,
\ee
where in the second step we have made use of the time evolution of $u$.
Since in the matter phase
\be
\eta_H(k) \, \sim \, k^{-1} \, ,
\ee
the factors of $k$ in (\ref{powerspec}) cancel and we indeed obtain a
scale-invariant spectrum.

Another way to reach this conclusion is to consider the equation of motion
of the curvature perturbation,
\eq{
 \ddot{\zeta_k} +3H \dot{\zeta_k} -\frac{k^2}{a^2}\zeta_k=0 \ ,
}
which has a solution
\eq{\label{zc}
 \zeta_k=A \frac{i c_s [1-ic_s k(\eta-\tilde{\eta_B})]}{\sqrt{2 c_s^3 k^3}(\eta-\tilde{\eta_B})^3} \exp [ic_s k (\eta-\tilde{\eta_B})] \ .
}
The constant factor $A$ is determined by the initial condition. Thus the spectrum of the curvature perturbation is scale-invariant in the contracting phase. The initial conditions yield
\begin{eqnarray}
A \equiv \frac{\sqrt{3(1+w)}H_B}{2M_p}(\frac{1+3w}{2}{\cal
H}_B)^{-\frac{3(1+w)}{1+3w}}~,
\end{eqnarray}
where $H_B$ is the Hubble scale, $\cal{H}_B$ is the conformal Hubble scale, and  the subscript ``$B$" denotes the momentum when the contracting phase ends.

Since the spectrum of fluctuations is scale-invariant and the fluctuations exit the Hubble radius
with an amplitude smaller than $1$, the amplitude of the fluctuations always remains perturbatively small.

\subsection{Ho\v{r}ava-Lifshitz Contracting Phase}

In this subsection we follow the fluctuation modes during the Ho\v{r}ava-Lifshitz contracting phase
between when they exit the Hubble radius with a vacuum spectrum until the end of the
contracting phase, when the higher derivative terms scaling as $a^{-4}$ in the action become
important. The terms scaling as $a^{-2}$ in the action which are induced by the spatial
curvature are negligible throughout since we are starting the evolution in the matter-dominated
contracting phase and therefore the curvature radius is larger than the wavelength we considered. Thus, it is a good approximation to first solve the curvature perturbation in flat space~($\bar{k}=0$),
and in the limit $\lambda \rightarrow 1$. In this limit, we can simplify the
expression for $\omega^2$ and obtain
\eq{ \label{omega1}
\omega^2=
\left(c^2 k^2-\frac{2}{\eta ^2}\right)+\frac{1}{12} c^2 k^2 \left(-24+c^2 k^2 \eta ^2-\frac{8 k^2}{\eta ^4 \Lambda  a_B^2}\right) (\lambda -1)
}
If $\lambda =1$, then
\eq{
\omega^2 \, =-\, \frac{2}{\eta ^2}+c^2k^2 \, .
}

For modes outside of the Hubble radius the $k^2$ term is negligible (note that Hubble radius
crossing corresponds to $k \eta = 1$). Hence, the correction terms to the mode equation
in Ho\v{r}ava-Lifshitz gravity compared to those in the Einstein theory are negligible in the
contracting phase. The same conclusion can be reached if we keep the leading terms due
to spatial curvature. Assuming that the universe is closed, the action for the background
becomes
\bea S^g_V &=& \int dt d^3x\sqrt{g}  N \frac{(1-3\lambda)c^3}{2\pi G}[\frac{-\frac{3}{2}+\frac{9(1-4\lambda)}{8(1-3\lambda)}}{\Lambda a^4}+\frac{3}{4(1-3\lambda)}\frac{1}{a^2}-\frac{3\Lambda}{8(1-3\lambda)}] \nonumber\\
&=& \int dt d^3x\sqrt{g}  N \frac{(1-3\lambda)c^3}{2\pi G} [\mathcal{O}(1)\frac{1}{\Lambda a^4}+\mathcal{O}(1)\frac{1}{a^2}+\mathcal{O}(1)\Lambda] \, .
\eea
In the limit $\Lambda a^2 \ll 1$ which is relevant in our case when we follow modes in a phase
in which the cosmological constant has a negligible effect, the $R^2$ terms dominate.
(The limit $\Lambda a^2 \gg 1$ would correspond to a phase during which the cosmological
constant is dominant. In this case the vacuum energy compensates the cosmological constant
term in Ho\v{r}ava-Lifshitz gravity.) In our case, Eq.(\ref{omega1}) can be expanded
to second order of $k$, yielding
\eq{
\omega^2=-\frac{2}{\eta ^2}+c^2(3-2\lambda)k^2 \ ,
}
from which it follows that the curvature perturbation is the same as in Eq.(\ref{zc}),
except that $c_s=c\sqrt{3-2\lambda}$.

To conclude, we see that the higher derivative terms in Ho\v{r}ava-Lifshitz gravity lead to
correction terms in the equation of motion for super-Hubble scale fluctuations.

\subsection{Ho\v{r}ava-Lifshitz Bouncing Phase}
\label{bouncing phase}

The contracting phase transits into the bouncing phase when the term in the action scaling
as $a^{-4}$ becomes important. Inspection of the Friedmann equation (\ref{friedmann}) shows
that this happens when
\be
T \, \sim \, T_0 ( \frac{M_{pl}}{\mu} )^2 \, .
\ee

During the bouncing phase the scale factor is given by (\ref{bounce}) with $y \sim \eta_B^{-2}$.
In this case, it follows from Eq.(\ref{omega}) that the mode frequency is given by
\bea
\omega^2 &=& [(-4+\pi ) ((-4+k^2) \kappa ^8 \mu ^4-2 \kappa ^4 \mu ^2 (2 \pi  y (5-6 \lambda -27 \lambda ^2+k^4 (1-4 \lambda +3 \lambda ^2)-2 k^2 (3-11 \lambda +6 \lambda ^2))\nonumber\\
&&
+(-5+k^2) \kappa ^4 \Lambda  \mu ^2) a_B^2+(-48 \pi ^2 y^2 (2+k^2 (-1+\lambda )) (-1+3 \lambda )^3\nonumber\\
&&
+4 \pi  y \kappa ^4 (k^4 (1-4 \lambda +3 \lambda ^2)-9 (-1+2 \lambda +3 \lambda ^2)-2 k^2 (4-15 \lambda +9 \lambda ^2)) \Lambda  \mu ^2+(-6+k^2) \kappa ^8 \Lambda ^2 \mu ^4) a_B^4)]/ \nonumber\\
&&
[64 (-3+k^2) y (-1+3 \lambda ) a_B^2 (\kappa ^4 (1+k^2 (-1+\lambda )-3 \lambda ) \mu ^2+(8 \pi  y (1-3 \lambda )^2+\kappa ^4 (-1-k^2 (-1+\lambda )+3 \lambda ) \Lambda  \mu ^2) a_B^2) \eta ^2]\nonumber \\
\eea
where we already assumed that $\bar{k}=1$.
Setting $\lambda = 1$ yields
\eq{
\omega^2=\frac{ (-4+\pi ) \left(2 c^2 \left(k^2-4 \right) +\left(-2 c^2 \left(k^2-6 K\right)-3 \pi  y\right) \Lambda  a_B^2\right)}{16 \left(k^2-3 \right) y \eta ^2 \Lambda  a_B^2} \, ,
}
and after some approximations one obtains
\begin{equation}
\omega^2= -\frac{\left(-4+k^2\right) (-4+\pi ) \kappa ^4 \mu ^2}{256 \left(-3+k^2\right) y a_B^2 \eta^2} \, .
\end{equation}

With this expression for the frequency, the solutions of the mode equation become
\begin{equation}
 \zeta = c_1' (\eta-\eta_B)^{(1-\sqrt{1+\frac{\left(-4+k^2\right) (-4+\pi ) \kappa ^4 \mu ^2}{64 \left(-3+k^2\right) y a_B^2 }})/2}+c_2'(\eta-\eta_B)^{(1+\sqrt{1+\frac{\left(-4+k^2\right) (-4+\pi ) \kappa ^4 \mu ^2}{64 \left(-3+k^2\right) y a_B^2 }})/2} \, .
\end{equation}

If we make the reasonable ``fast bounce assumption"  $\frac{y a_B^2}{\mu^2 \kappa^4}\gg 1 $
then we find that for IR modes ( i.e. modes with $k\ll 1$)
the value of $\zeta$ is almost unchanged between the
beginning and end of the bounce phase, i.e.
\begin{equation}
 |\zeta_e| \, \simeq \, |\zeta_c| \, ,
\end{equation}
where $\zeta_c (\zeta_e)$ denotes the value of $\zeta$ at the end of the contracting phase
and beginning of the expanding phase, respectively.

Note, in particular, that - as follows from the equation of the curvature perturbation in the bouncing phase - there is no singularity or instability in the solution for $\zeta$. Thus, we conclude
that the fluctuations pass through the bounce without singularity and without change in the
spectrum.

\subsection{Expanding Phase}

In the expanding phase, the mode equation for $\zeta$ has two fundamental solutions.
The dominant mode is constant in time on super-Hubble scales, the second one the decaying.
Thus, the spectrum of $\zeta$ on super-Hubble scales at late times is the same one as
emerges after the bounce at the beginning of the expanding phase.

\subsection{Matching Condtion}

As we have seen, there are three phases in the matter bounce senario. In each
phase we have derived approximate analytical solutions of the mode equations.
All that remains is to match them correctly. The matching conditions for
cosmological perturbations across a space-like slice were discussed by Hwang and
Vishniac \cite{HV} and by Deruelle and Mukhanov \cite{DM}. These works show
that $\zeta$ and $\Phi$ must be continuous across the transition surface.

As stressed in \cite{Durrer}, these matching conditions for fluctuations are only
applicable if the background satisfies the continuity of both the induced metric
and the extrinsic curvature on the
matching surface . If one were to match across a singular
transition between contraction and expansion as was done \cite{done} in four
dimensional toy models of the Ekpyrotic \cite{Ekp} scenario, then the background
does not satisfy the matching conditions and hence the applicability of the matching
conditions to the fluctuations is questionable. However, in a non-singular bouncing
cosmology such as the one we are considering here we can apply the matching
conditions consistently at the transition between the contracting matter phase and
bounce phase, and between the bounce phase and the expanding matter phase.
This procedure has already been applied in the case of the nonsingular mirage
cosmology bounce of \cite{BFS}, the higher derivative gravity bounce \cite{ABB}, and
in the quintom and Lee-Wick bounces \cite{recent}.

Matching between the contacting phase and the bouncing phase implies that the
spectrum of $\zeta$ at the beginning of the bounce phase is the same as it is
at the end of the contracting phase, namely scale-invariant. Since the mode
functions are the same at the beginning and end of the bounce phase it follows
that the spectrum is scale-invariant at the end of the bounce phase. Matching
at the transition between the bounce phase and the expanding phase preserves
the scale-invariance of the spectrum. Hence, we conclude that the spectrum of
cosmological perturbations is scale-invariant at late times.

More specifically, the values of the mode functions in the expanding phase
are given by
\eq{
\zeta_e = \frac{\sqrt{3}c_s H_B }{2 M_P \sqrt{2c_s^3 k^3}} \exp (ic_s k\eta) \, ,
}
and hence the spectrum of curvature perturbation is
\eq{
P_{\zeta}=\frac{k^3}{2\pi^2} |\zeta_c|^2 =  \frac{3 H_B^2}{4\pi^2 c_s M_p^2}
}
where $c_s =\sqrt{3-2\lambda}$, and $H_B$ is the value of $|H|$ at the end
of the contracting phase (the maximal value of $|H|$.

\section{Conclusions}

We have studied the evolution of linear cosmological perturbations in a bouncing
Ho\v{r}ava-Lifshitz cosmology. We have seen that at linear order in perturbation
theory there are no extra dynamical degrees of freedom, the same conclusion as
was reached in an expanding Ho\v{r}ava-Lifsthitz cosmology \cite{us}.

The equations of motion for the fluctuations contain a singularity at the bounce point.
We have seen, however, that the solutions are non-singular and thus can be
smoothly extended from the contracting to the expanding phase. We have
derived approximate solutions of the equations of motion in the contracting and
bounce phases. We have seen that the extra terms in the Ho\v{r}ava-Lifshitz action
have a negligible effect on the evolution of fluctuations on super-Hubble scales.
Thus, an initial vacuum spectrum of sub-Hubble fluctuations in the far past evolves
into a scale-invariant spectrum of curvature fluctuations on super-Hubble scales
at the end of the contracting phase. Because of the smooth matching of the
fluctuations between the contracting phase and the bounce phase, and between
the bounce phase and the expanding phase, and because of the fact that modes
which are super-Hubble at the end of the contracting phase hardly change between
the beginning and end of the bounce phase, the scale-invariance of the spectrum
of cosmological perturbations is preserved during the bounce, as initially
conjectured in \cite{RHB}.

We have seen that initial vacuum fluctuations lead to a power spectrum which
is perturbatively small throughout the bounce phase. However, if one were
to arrange the value of the spatial curvature and the cosmological constant
such that a cyclic background would result, the non-trivial evolution of
fluctuations on super-Hubble scales in the contracting phase would destroy
the cyclicity of the evolution \cite{RHBcyclic}. The fluctuations would no longer
be perturbatively small during the second bounce. Thus, one should not
consider values of the parameters in the Ho\v{r}ava-Lifshitz action which would
lead to a cyclic background.

\acknowledgements

We thank for Yifu Cai for useful discussion.
This research is supported at McGill by a NSERC Discovery Grant and
by funds from the Canada Research Chairs program. YW is supported in
part by a Institute of Particle Physics postdoctoral fellowship, and
by funds from McGill University. XG was supported by the NSFC grant
No.10535060/A050207, a NSFC group grant No.10821504 and Ministry of
Science and Technology 973 program under grant No.2007CB815401.


\appendix

\section{Various Coefficients}

\subsection{Coefficients in (\protect\ref{S2_psi_Q})}{\label{appsec_coeff_S2_1}}

The coefficients which appear in Eq. ( \ref{S2_psi_Q} ) are
\ea{
    c_{\psi} &= \frac{4 \left(3 \bar{k}+a^2 \Delta \right) (-1+3 \lambda ) \dop_0^2}{8 H^2 \left(3 \bar{k}+a^2 \Delta \right) (-1+3 \lambda )+\kappa ^2 \left(-2 \bar{k}+a^2 \Delta  (-1+\lambda )\right) \dop_0^2}  \,,
}

\ea{
    c_{Q} &=  \frac{4 H^2 \left(3 \bar{k}+a^2 \Delta \right) (-1+3 \lambda )}{8 H^2 \left(3 \bar{k}+a^2 \Delta \right) (-1+3 \lambda )+\kappa ^2 \left(-2 \bar{k}+a^2 \Delta  (-1+\lambda )\right) \dop_0^2} \,,
}

\ea{
    c_{c} &= \frac{8 H \left(3 \bar{k}+a^2 \Delta \right) (-1+3 \lambda ) \dop_0}{8 H^2 \left(3 \bar{k}+a^2 \Delta \right) (-1+3 \lambda )+\kappa ^2 \left(-2 \bar{k}+a^2 \Delta  (-1+\lambda )\right) \dop_0^2}  \,,
}

\ea{
    f_{\psi} &=  -\frac{ 4  H}{a^4 \kappa ^2 \left(8 H^2 \left(3 \bar{k}+a^2 \Delta \right) (-1+3 \lambda )+\kappa ^2 \left(-2 \bar{k}+a^2 \Delta  (-1+\lambda )\right) \dop_0^2\right)} \\
    &\qquad \times \lrsb{ \left(3 \bar{k}+a^2 \Delta \right) \left(24 a^4 H^2 (1-3 \lambda )^2+\left(3 \bar{k}+a^2 \Delta \right) \kappa ^4 \left(\bar{k}-a^2 \Lambda \right) \mu ^2\right)+3 a^4 \kappa ^2 \left(-2 \bar{k}+a^2 \Delta  (\lambda-1 )\right) (3 \lambda-1 ) \dop_0^2 }\,,
}

\ea{
    f_{Q} &= \frac{ \kappa ^2 \dop_0 \left(a^2 H \Delta  (1-3 \lambda ) \dop_0+\left(2 \bar{k}-a^2 \Delta  (-1+\lambda )\right) V'\right)}{8 H^2 \left(3 \bar{k}+a^2 \Delta \right) (-1+3 \lambda )+\kappa ^2 \left(-2 \bar{k}+a^2 \Delta  (-1+\lambda )\right) \dop_0^2}  \,,
}

\ea{
    f_{c} &= \frac{ (-1+3 \lambda ) \left(-a^2 \Delta  \kappa ^2 \dop_0^3+8 H \left(3 \bar{k}+a^2 \Delta \right) V'\right)}{8 H^2 \left(3 \bar{k}+a^2 \Delta \right) (-1+3 \lambda )+\kappa ^2 \left(-2 \bar{k}+a^2 \Delta  (-1+\lambda )\right) \dop_0^2}  \,,
}

\ea{
    \tilde{f}_{c} &=  -\frac{\dop_0 }{2 a^4 (-1+3 \lambda ) \left(8 H^2 \left(3 \bar{k}+a^2 \Delta \right) (-1+3 \lambda )+\kappa ^2 \left(-2 \bar{k}+a^2 \Delta  (\lambda-1 )\right) \dop_0^2\right)} \\
    &\qquad  \times \lrsb{ \left(3 \bar{k}+a^2 \Delta \right) \left(48 a^4 H^2 (1-3 \lambda )^2+\kappa ^4 \left(2 \bar{k}-a^2 \Delta  (\lambda -1)\right) \left(\bar{k}-a^2 \Lambda \right) \mu ^2\right)+6 a^4 \kappa ^2 \left(-2 \bar{k}+a^2 \Delta  (\lambda-1 )\right) (3 \lambda -1) \dop_0^2 } \,,
}

\ea{
    \omega_{\psi} &=  \frac{1}{8 a^8 (\kappa -3 \kappa  \lambda )^2 \left(8 H^2 \left(3 \bar{k}+a^2 \Delta \right) (-1+3 \lambda )+\kappa ^2 \left(-2 \bar{k}+a^2 \Delta  (-1+\lambda )\right) \dop_0^2\right)} \\
    &\qquad \times \bigg\{  \left(3 \bar{k}+a^2 \Delta \right) \Big[ -1152 a^8 H^4 (1-3 \lambda )^4-16 a^6 H^2 \kappa ^4 (1-3 \lambda )^2 \left(\bar{k} (\Delta  (-4+3 \lambda )-6 \Lambda )+a^2 \Delta  (\Delta  (-1+\lambda )-\Lambda )\right) \mu ^2  \\
    &\qquad\qquad  +\left(3 \bar{k}+a^2 \Delta \right) \kappa ^8 \left(2 \bar{k}-a^2 \Delta  (-1+\lambda )\right) \left(\bar{k}-a^2 \Lambda \right)^2 \mu ^4 \Big] \\
    &\qquad - 2 a^6 \kappa ^2 (-1+3 \lambda ) \dop_0^2 \Big[ 24 a^2 H^2 (1-3 \lambda )^2 \left(-12 \bar{k}+a^2 \Delta  (-5+3 \lambda )\right) \\
    &\qquad\qquad\qquad\qquad -\kappa ^4 \left(2 \bar{k}-a^2 \Delta  (-1+\lambda )\right) \left(\bar{k} (-4 \Delta +3 \Delta  \lambda -6 \Lambda )+a^2 \Delta  (\Delta  (-1+\lambda )-\Lambda )\right) \mu ^2 \\
    &\qquad\qquad\qquad\qquad -6 a^2 \kappa ^2 \left(-2 \bar{k}+a^2 \Delta  (-1+\lambda )\right) (-1+3 \lambda ) \dop_0^2 \Big] \bigg\} \,,
}

\ea{
    \omega_{Q} &= \frac{1}{64 H^2 \left(3 \bar{k}+a^2 \Delta \right) (-1+3 \lambda )-8 \kappa ^2 \left(2 \bar{k}-a^2 \Delta  (-1+\lambda )\right) \dop_0^2} \\
    &\qquad \times \bigg\{  -192 H^2 \bar{k} \Delta ^3 g_3-64 a^2 H^2 \Delta ^4 g_3+576 H^2 \bar{k} \Delta ^3 \lambda  g_3+192 a^2 H^2 \Delta ^4 \lambda  g_3+12 a^2 H^2 \Delta  \kappa ^2 \dop_0^2 \\
    &\qquad\qquad -36 a^2 H^2 \Delta  \kappa ^2 \lambda  \dop_0^2-16 \bar{k} \Delta ^3 \kappa ^2 g_3 \dop_0^2-8 a^2 \Delta ^4 \kappa ^2 g_3 \dop_0^2+8 a^2 \Delta ^4 \kappa ^2 \lambda  g_3 \dop_0^2+a^2 \Delta  \kappa ^4 \dop_0^4 \\
    &\qquad\qquad + 8 \Delta  g_1 \left(-8 H^2 \left(3 \bar{k}+a^2 \Delta \right) (-1+3 \lambda )+\kappa ^2 \left(2 \bar{k}-a^2 \Delta  (-1+\lambda )\right) \dop_0^2\right) \\
    &\qquad\qquad +8 \Delta ^2 g_2 \left(8 H^2 \left(3 \bar{k}+a^2 \Delta \right) (-1+3 \lambda )+\kappa ^2 \left(-2 \bar{k}+a^2 \Delta  (-1+\lambda )\right) \dop_0^2\right) \\
    &\qquad\qquad +8 a^2 H \Delta  \kappa ^2 \dop_0 V'-24 a^2 H \Delta  \kappa ^2 \lambda  \dop_0 V'+8 \bar{k} \kappa ^2 \left(V'\right)^2+4 a^2 \Delta  \kappa ^2 \left(V'\right)^2-4 a^2 \Delta  \kappa ^2 \lambda  \left(V'\right)^2 \\
    &\qquad\qquad +4 \left(-8 H^2 \left(3 \bar{k}+a^2 \Delta \right) (-1+3 \lambda )+\kappa ^2 \left(2 \bar{k}-a^2 \Delta  (-1+\lambda )\right) \dop_0^2\right) V'' \bigg\} \,,
}

\ea{
    \omega_{c} &= \frac{1}{2 a^4 (-1+3 \lambda ) \left(8 H^2 \left(3 \bar{k}+a^2 \Delta \right) (-1+3 \lambda )+\kappa ^2 \left(-2 \bar{k}+a^2 \Delta  (-1+\lambda )\right) \dop_0^2\right)} \\
    &\qquad \times \bigg\{ a^2 H \Delta  \left(3 \bar{k}+a^2 \Delta \right) \kappa ^4 (-1+3 \lambda ) \left(\bar{k}-a^2 \Lambda \right) \mu ^2 \dop_0 \\
    &\qquad\qquad + \Big[ \left(3 \bar{k}+a^2 \Delta \right) \left(48 a^4 H^2 (1-3 \lambda )^2-\kappa ^4 \left(2 \bar{k}-a^2 \Delta  (-1+\lambda )\right) \left(\bar{k}-a^2 \Lambda \right) \mu ^2\right) \\
    &\qquad\qquad\qquad\qquad + 6 a^4 \kappa ^2 \left(-2 \bar{k}+a^2 \Delta  (-1+\lambda )\right) (-1+3 \lambda ) \dop_0^2 \Big] V' \bigg\} \,,
}

\subsection{Coefficients in
(\protect\ref{S2_zeta_psi})}\label{appsec_coeff_S2_2}

The coefficients which appear in Eq. (\ref{S2_zeta_psi}) are
\ea{
    c_{\zeta} &= \frac{4 \left(3 \bar{k}+a^2 \Delta \right) (-1+3 \lambda ) \dop_0^2}{8 H^2 \left(3 \bar{k}+a^2 \Delta \right) (-1+3 \lambda )+\kappa ^2 \left(-2 \bar{k}+a^2 \Delta  (-1+\lambda )\right) \dop_0^2}  \,,
}
    \ea{
    f_{\zeta} &= \frac{-\dop_0}{a^4 H^2 (-1+3 \lambda ) \left(8 H^2 \left(3 \bar{k}+a^2 \Delta \right) (-1+3 \lambda )+\kappa ^2 \left(-2 \bar{k}+a^2 \Delta  (-1+\lambda )\right) \dop_0^2\right)} \\
    &\qquad \times \Big\{  H \left(3 \bar{k}+a^2 \Delta \right) \left[24 a^4 H^2 (1-3 \lambda )^2+\bar{k} \kappa ^4 \left(\bar{k}-a^2 \Lambda \right) \mu ^2\right] \dop_0   \\
    &\qquad\qquad +3 a^4 H \kappa ^2 \left(-2 \bar{k}+a^2 \Delta  (-1+\lambda )\right) (-1+3 \lambda ) \dop_0^3+8 a^4 H^2 \left(3 \bar{k}+a^2 \Delta \right) (1-3 \lambda )^2 V'  \\
    &\qquad\qquad +a^4 \kappa ^2 \left(-2 \bar{k}+a^2 \Delta  (-1+\lambda )\right) (-1+3 \lambda ) \dop_0^2 V' \Big\} \,,
}
    \ea{
    \bar{f}_{c} &=  \frac{\Delta  \left(3 \bar{k}+a^2 \Delta \right) \kappa ^4 (-1+\lambda ) \left(-\bar{k}+a^2 \Lambda \right) \mu ^2 \dop_0^2}{2 a^2 H (-1+3 \lambda ) \left(8 H^2 \left(3 \bar{k}+a^2 \Delta \right) (-1+3 \lambda )+\kappa ^2 \left(-2 \bar{k}+a^2 \Delta  (-1+\lambda )\right) \dop_0^2\right)} \,,
}

\begin{eqnarray*}
\omega_{\zeta} & = & \frac{1}{16a^{8}H^{3}(-1+3\lambda)^{3}\left(8H^{2}\left(k^{2}-3\bar{k}\right)(-1+3\lambda)+\kappa^{2}\left(2\bar{k}+k^{2}(-1+\lambda)\right)\dop_{0}^{2}\right)}\\
 &  & \times\left(-6a^{8}H\kappa^{4}\left(2\bar{k}+k^{2}(-1+\lambda)\right)(1-3\lambda)^{2}\dop_{0}^{6}+\right.\\
 &  & 16a^{4}\left(k^{2}-3\bar{k}\right)(H-3H\lambda)^{2}\left(24a^{4}H^{2}(1-3\lambda)^{2}+\bar{k}\kappa^{4}\left(\bar{k}-a^{2}\Lambda\right)\mu^{2}\right)\dop_{0}V'+\\
 &  & 2a^{4}\kappa^{2}(-1+3\lambda)\left(8a^{4}(H-3H\lambda)^{2}\left(12\bar{k}+k^{2}(-5+3\lambda)\right)+\bar{k}\kappa^{4}\left(2\bar{k}+k^{2}(-1+\lambda)\right)\left(\bar{k}-a^{2}\Lambda\right)\mu^{2}\right)\dop_{0}^{3}V'-\\
 &  & 4a^{8}\kappa^{4}\left(2\bar{k}+k^{2}(-1+\lambda)\right)(1-3\lambda)^{2}\dop_{0}^{5}V'+64a^{8}H^{3}\left(k^{2}-3\bar{k}\right)(1-3\lambda)^{4}\left(V'\right)^{2}+\\
 &  & 2a^{2}H\kappa^{2}(-1+3\lambda)\dop_{0}^{4}\left(12a^{6}(H-3H\lambda)^{2}\left(12\bar{k}+k^{2}(-5+3\lambda)\right)+3a^{2}\bar{k}\kappa^{4}\left(2\bar{k}+k^{2}(-1+\lambda)\right)\left(\bar{k}-a^{2}\Lambda\right)\mu^{2}-\right.\\
 &  & \left.4\left(2\bar{k}+k^{2}(-1+\lambda)\right)(1-3\lambda)^{2}\left(-2a^{4}k^{2}g_{1}-2a^{2}k^{4}g_{2}+2k^{6}g_{3}+a^{6}V''\right)\right)+\\
 &  & H\dop_{0}^{2}\left(\left(k^{2}-3\bar{k}\right)\left(24a^{4}H^{2}(1-3\lambda)^{2}+\bar{k}\kappa^{4}\left(\bar{k}-a^{2}\Lambda\right)\mu^{2}\right)^{2}+128a^{6}H^{2}k^{2}\left(k^{2}-3\bar{k}\right)(1-3\lambda)^{4}g_{1}+\right.\\
 &  & 128a^{4}H^{2}k^{4}\left(k^{2}-3\bar{k}\right)(1-3\lambda)^{4}g_{2}+8a^{2}(-1+3\lambda)^{3}\\
 &  & \left.\left.\left(-16H^{2}k^{6}\left(k^{2}-3\bar{k}\right)(-1+3\lambda)g_{3}+a^{6}\left(\kappa^{2}\left(2\bar{k}+k^{2}(-1+\lambda)\right)\left(V'\right)^{2}-8H^{2}\left(k^{2}-3\bar{k}\right)(-1+3\lambda)V''\right)\right)\right)\right)\end{eqnarray*}

\begin{eqnarray*}
\bar{\omega}_{c} & = & \frac{1}{16a^{8}H^{3}(-1+3\lambda)^{3}\left(8H^{2}\left(k^{2}-3\bar{k}\right)(-1+3\lambda)+\kappa^{2}\left(2\bar{k}+k^{2}(-1+\lambda)\right)\dop_{0}^{2}\right)}\\
 &  & \times\left(-2a^{2}Hk^{2}\kappa^{2}\left(2\bar{k}+k^{2}(-1+\lambda)\right)(-1+3\lambda)\left(a^{2}\kappa^{4}\left(-\bar{k}+a^{2}\Lambda\right)\mu^{2}-16(1-3\lambda)^{2}\left(a^{4}g_{1}+a^{2}k^{2}g_{2}-k^{4}g_{3}\right)\right)\dop_{0}^{4}+\right.\\
 &  & 32a^{4}\left(k^{2}-3\bar{k}\right)(H-3H\lambda)^{2}\left(12a^{4}H^{2}(1-3\lambda)^{2}+\bar{k}\kappa^{4}\left(\bar{k}-a^{2}\Lambda\right)\mu^{2}-8a^{4}(1-3\lambda)^{2}\dot{H}\right)\dop_{0}V'+\\
 &  & 4a^{4}\kappa^{2}(-1+3\lambda)\left(4a^{4}(H-3H\lambda)^{2}\left(18\bar{k}+k^{2}(-7+3\lambda)\right)+\right.\\
 &  & \left.\bar{k}\kappa^{4}\left(2\bar{k}+k^{2}(-1+\lambda)\right)\left(\bar{k}-a^{2}\Lambda\right)\mu^{2}-8a^{4}\left(2\bar{k}+k^{2}(-1+\lambda)\right)(1-3\lambda)^{2}\dot{H}\right)\dop_{0}^{3}V'-\\
 &  & 8a^{8}\kappa^{4}\left(2\bar{k}+k^{2}(-1+\lambda)\right)(1-3\lambda)^{2}\dop_{0}^{5}V'+128a^{8}H^{3}\left(k^{2}-3\bar{k}\right)(1-3\lambda)^{4}V'\left(\tilde{\phi}_{0}+V'\right)+\\
 &  & H\dop_{0}^{2}\left(-k^{2}\left(k^{2}-3\bar{k}\right)\kappa^{4}\left(\bar{k}-a^{2}\Lambda\right)\mu^{2}\left(-16a^{4}H^{2}(1-3\lambda)^{2}+\bar{k}\kappa^{4}(-1+\lambda)\left(\bar{k}-a^{2}\Lambda\right)\mu^{2}\right)+\right.\\
 &  & 16a^{2}(-1+3\lambda)^{3}\left(16H^{2}k^{2}\left(k^{2}-3\bar{k}\right)(-1+3\lambda)\left(a^{4}g_{1}+a^{2}k^{2}g_{2}-k^{4}g_{3}\right)+\right.\\
 &  & \left.\left.\left.a^{6}\kappa^{2}\left(2\bar{k}+k^{2}(-1+\lambda)\right)\tilde{\phi}_{0}V'+a^{6}\kappa^{2}\left(2\bar{k}+k^{2}(-1+\lambda)\right)\left(V'\right)^{2}\right)\right)\right)\end{eqnarray*}

\begin{eqnarray*}
\bar{\omega}_{\psi} & = & \frac{1}{16a^{8}H^{3}\kappa^{2}(-1+3\lambda)^{3}}\\
 &  & \left(4a^{4}(1-3\lambda)^{2}\dot{H}\left(24a^{4}H^{3}(1-3\lambda)^{2}+H\left(k^{2}-3\bar{k}\right)\kappa^{4}\left(\bar{k}-a^{2}\Lambda\right)\mu^{2}+2a^{4}\kappa^{2}(-1+3\lambda)\dop_{0}\left(3H\dop_{0}-2V'\right)\right)+\right.\\
 &  & \left(\kappa^{2}\left(-2a^{2}H\kappa^{2}(-1+3\lambda)\left(12a^{6}(H-3H\lambda)^{2}\left(16\bar{k}+k^{2}(-7+5\lambda)\right)+a^{2}\left(k^{2}-3\bar{k}\right)\kappa^{4}\left(2\bar{k}+k^{2}(-1+\lambda)\right)\right.\right.\right.\\
 &  & \left.\left(-\bar{k}+a^{2}\Lambda\right)\mu^{2}-8\left(2\bar{k}+k^{2}(-1+\lambda)\right)(k-3k\lambda)^{2}\left(a^{4}g_{1}+a^{2}k^{2}g_{2}-k^{4}g_{3}\right)\right)\dop_{0}^{4}+\\
 &  & 6a^{8}H\kappa^{4}\left(2\bar{k}+k^{2}(-1+\lambda)\right)(1-3\lambda)^{2}\dop_{0}^{6}-4a^{8}\kappa^{4}\left(2\bar{k}+k^{2}(-1+\lambda)\right)(1-3\lambda)^{2}\dop_{0}^{5}V'-\\
 &  & 2a^{4}\kappa^{2}(-1+3\lambda)\dop_{0}^{3}\left(24a^{4}\left(2\bar{k}+k^{2}(-1+\lambda)\right)(H-3H\lambda)^{2}\tilde{\phi}_{0}+\right.\\
 &  & \left.\left(4a^{4}H^{2}(1-3\lambda)^{2}\left(k^{2}-6\bar{k}+3k^{2}\lambda\right)-\bar{k}\kappa^{4}\left(2\bar{k}+k^{2}(-1+\lambda)\right)\left(\bar{k}-a^{2}\Lambda\right)\mu^{2}\right)V'\right)-\\
 &  & 16a^{4}\left(k^{2}-3\bar{k}\right)(H-3H\lambda)^{2}\dop_{0}\left(24a^{4}H^{2}(1-3\lambda)^{2}\tilde{\phi}_{0}+\left(12a^{4}H^{2}(1-3\lambda)^{2}+\bar{k}\kappa^{4}\left(-\bar{k}+a^{2}\Lambda\right)\mu^{2}\right)V'\right)+\\
 &  & 2H^{3}\left(k^{2}-3\bar{k}\right)(-1+3\lambda)\left(16a^{4}H^{2}\kappa^{2}(1-3\lambda)^{2}\left(-k^{4}(-1+\lambda)+3k^{2}\bar{k}(-1+\lambda)-3\bar{k}\left(\bar{k}-a^{2}\Lambda\right)\right)\mu^{2}-\right.\\
 &  & \left.\left(k^{2}-3\bar{k}\right)\kappa^{6}\left(2\bar{k}+k^{2}(-1+\lambda)\right)\left(\bar{k}-a^{2}\Lambda\right)^{2}\mu^{4}+32a^{8}(-1+3\lambda)^{3}V'\left(\tilde{\phi}_{0}+V'\right)\right)+\\
 &  & H\dop_{0}^{2}\left(-960a^{8}H^{4}\left(k^{2}-3\bar{k}\right)(1-3\lambda)^{4}-4a^{4}H^{2}\kappa^{4}(1-3\lambda)^{2}\left(k^{6}(-1+\lambda)^{2}-30\bar{k}^{2}\left(\bar{k}-a^{2}\Lambda\right)+\right.\right.\\
 &  & \left.k^{4}\left(\bar{k}(-9+(8-3\lambda)\lambda)+4a^{2}\Lambda\right)-3k^{2}\bar{k}\left(\bar{k}(-9+\lambda)+a^{2}(7+\lambda)\Lambda\right)\right)\mu^{2}-\\
 &  & \left(k^{2}-3\bar{k}\right)\bar{k}\kappa^{8}\left(\bar{k}+k^{2}(-1+\lambda)\right)\left(\bar{k}-a^{2}\Lambda\right)^{2}\mu^{4}+8a^{2}(-1+3\lambda)^{3}\left(16H^{2}k^{2}\left(k^{2}-3\bar{k}\right)(-1+3\lambda)\right.\\
 &  & \left.\left.\left.\left.\left.\left(a^{4}g_{1}+a^{2}k^{2}g_{2}-k^{4}g_{3}\right)+a^{6}\kappa^{2}\left(2\bar{k}+k^{2}(-1+\lambda)\right)\tilde{\phi}_{0}V'+a^{6}\kappa^{2}\left(2\bar{k}+k^{2}(-1+\lambda)\right)\left(V'\right)^{2}\right)\right)\right)\right)\right/\\
 &  & \left.\left(8H^{2}\left(k^{2}-3\bar{k}\right)(-1+3\lambda)+\kappa^{2}\left(2\bar{k}+k^{2}(-1+\lambda)\right)\dop_{0}^{2}\right)\right)\end{eqnarray*}

These expressions simplify dramatically if one sets
$\lambda =1$ and $\bar{k}=0$. Setting, in addition, $w = 0$ as is appropriate
for the contracting phase $w=0$, one gets
\begin{eqnarray*}
c_\zeta &=&\frac{6}{\kappa ^2} \\
f_\zeta&=&-\frac{36}{\eta ^3 \kappa ^2 a_B}\\
\bar{f}_{c} & = & 0\\
\bar{\omega}_{c} & = & -\frac{6 \left(9+k^2 \eta ^2\right)}{\eta ^6 \kappa ^2 a_B^2} \\
\bar{\omega}_{c} & = &\frac{3 k^2 \left(-\kappa ^4 \Lambda  M^2+\frac{64 g_1}{\eta ^4 a_B^2}\right)}{8 \kappa ^2}\\
\bar{\omega}_{\psi} & = & \frac{3 k^2 \left(-\kappa ^4 \Lambda  M^2+\frac{64 g_1}{\eta ^4 a_B^2}\right)}{16 \kappa ^2}
\end{eqnarray*}
where $M=\mu/a$

Inserting these coefficients into the expression for the frequency $\omega$ one obtains
\begin{equation}
\omega^2 \, = \, -\frac{2}{\eta ^2} + c^2 k^2 \, .
\end{equation}

If the condition $\lambda = 1$ is slightly relaxed and one  expands to first order in
$\lambda-1$ then one gets
\begin{equation}
\omega^2 \, = \,
\left(c^2 k^2-\frac{2}{\eta ^2}\right)+\frac{1}{12} c^2 k^2 \left(-24+c^2 k^2 \eta ^2-\frac{8 k^2}{\eta ^4 \Lambda  a_B^2}\right) (\lambda -1) \, .
\end{equation}

\section{Useful Formulae}

\subsection{Preliminaries}

The connection for the perturbed metric (\ref{metric_pert_form}) is
(up to second-order):
    \ea{
        \Gamma^i_{ij(0)} &\equiv  - \frac{\bar{k}}{2 \lrp{1 + \frac{\bar{k}r^2}{4}}} \lrp{ x^j \delta_{ik} + x^k \delta_{ij} - x^i \delta_{jk}
    } \,,\\
        \Gamma^i_{ij(1)} &\equiv - \lrp{ \rd_j \psi \delta_{ik} + \rd_k\psi \delta_{ij} - \rd_i\psi \delta_{jk}
    }  \,,\\
    \Gamma^i_{ij(2)} &\equiv -2 \lrp{ \psi\rd_j \psi \delta_{ik} + \psi\rd_k\psi \delta_{ij} - \psi\rd_i\psi \delta_{jk}
    }  \,.
    }

The Laplacian for the metric (\ref{metric_pert_form}) including fluctuations is
$\Delta \equiv \Delta_{(0)} + \Delta_{(1)} + \dots$, with\ea{
        \Delta_{(0)} &\equiv g^{ij}_{(0)} \rd_i\rd_j - g^{ij}_{(0)} \Gamma^k_{ij(0)} \rd_k  \,,\\
        \Delta_{(1)} &\equiv  g^{ij}_{(1)} \rd_i\rd_j - g^{ij}_{(0)}  \Gamma^k_{ij(1)} \rd_k
         - g^{ij}_{(1)} \Gamma^k_{ij(0)}
        \rd_k \,,
    }
where $g_{(0)_{ij}} \equiv \bar{g}_{ij}$. For our purpose, we do not
need higher-order terms such as $\Delta_{(2)}$ etc. Obviously,
$\Delta_{(0)}$ is just the background Laplacian (acting on scalar
fields) and $\Delta_{(1)}$ is the first-order backreaction on the
Laplacian due to the fluctuation $\psi$.

\subsection{Expansions of Various Quantities}

{F}or the metric (\ref{metric_pert_form}), we have
    \ea{
        \mathcal{E} &\equiv E_{ij}E^{ij} - \lambda E^2 \\
        &= 3 (1-3\lambda)\lrp{H - \frac{\dot{\psi}}{1-2\psi}}^2 - 2 (1-3\lambda)\lrp{H - \frac{\dot{\psi}}{1-2\psi}}\Delta
        B + \lrsb{ \nabla_i\nabla_j B \, \nabla^i\nabla^j B - \lambda \lrp{\Delta B}^2
        } \,,
    }
and
    \ea{
        R_{ij} = \frac{1}{1-2\psi} \lrsb{ \lrp{\frac{2\bar{k}}{a^2} + \Delta\psi + \frac{3\rd_k\psi \rd^k\psi}{1-2\psi}} g_{ij} + \lrp{ \nabla_i\nabla_j\psi + \frac{\rd_i\psi \rd_j\psi}{1-2\psi} }
        } \,,
    }
with $\rd_i\psi \rd^i\psi \equiv g^{ij}\rd_i\psi \rd_j\psi$. The
above results are exact.

Now it is straightforward  to expand various quantities up to
second-order in perturbations. In this appendix we simply collect
the final results.
\itm{
    \item $\mathcal{E}\equiv E_{ij}E^{ij} - \lambda E^2$

Denote $\mathcal{E} \equiv E_{ij}E^{ij} - \lambda E^2$ as a
shorthand. Then, to second-order in perturbations, we have
    \ea{
        \mathcal{E}_{(0)} &\equiv  3(1-3\lambda)H^2 \,,\\
        \mathcal{E}_{(1)} &\equiv  -2H (1-3\lambda) \lrsb{3 \dot{\psi} + \Delta_{(0)}B } \,,\\
        \mathcal{E}_{(2)} &\equiv  (1-3\lambda) \lrsb{ 3\dot{\psi}^2 +2\dot{\psi} \Delta_{(0)}B -12 H\dot{\psi }\psi -6H \psi \Delta_{(0)}B } + \lrsb{ (1 - \lambda)
        \lrp{\Delta_{(0)}B}^2 + \frac{2\bar{k}}{a^2}B \Delta_{(0)}B
        } \,.
    }

    \item $R$
    \ea{
        R_{(0)} &\equiv  \frac{6\bar{k}}{a^2} \,,\\
        R_{(1)} &\equiv  \frac{4}{a^2} \lrp{ a^2 \Delta_{(0)} \psi + 3\bar{k}\psi
        } \,,\\
        R_{(2)} &\equiv  2 \lrp{ 5 \,\psi\, \Delta_{(0)}\psi + \frac{12\bar{k}}{a^2}\psi^2
        }  \,.
    }

    \item $R^2$    \ea{
        \lrp{R^2}_{(0)} &\equiv \frac{36 \bar{k}^2}{a^4}  \,,\\
        \lrp{R^2}_{(1)} &\equiv \frac{48 \bar{k}}{a^4} \lrp{ a^2 \Delta_{(0)} \psi + 3 \bar{k} \psi }  \,,\\
        \lrp{R^2}_{(2)} &\equiv  8 \lrsb{ 2 \lrp{\Delta_{(0)} \psi}^2+ \frac{54 \bar{k}^2}{a^4} \psi ^2+ \frac{27 \bar{k}}{a^2}  \psi \Delta_{(0)} \psi
        } \,.
    }

    \item $R_{ij}R^{ij}$    \ea{
        \lrp{ R_{ij}R^{ij} }_{(0)} &\equiv \frac{12\bar{k}^2}{a^4}  \,,\\
        \lrp{ R_{ij}R^{ij} }_{(1)} &\equiv  \frac{16\bar{k}}{a^4}
        \lrp{ a^2 \Delta_{(0)}\psi +3\bar{k}\psi } \,,\\
        \lrp{ R_{ij}R^{ij} }_{(2)} &\equiv  6 \lrp{\Delta_{(0)}\psi }^2   + \frac{74\bar{k}}{a^2} \psi \Delta_{(0)}\psi  + \frac{144\bar{k}^2}{a^4} \psi^2
        \,.
    }
}


%
%



\begin{thebibliography}{99}

\bibitem{Horava0}
P.~Horava,
  ``Membranes at Quantum Criticality,''
  JHEP {\bf 0903}, 020 (2009)
  [arXiv:0812.4287 [hep-th]].

\bibitem{Horava1}
P.~Ho\v{r}ava,
  ``Quantum Gravity at a Lifshitz Point,''
  Phys. Rev. {\bf D 79}, 084008 (2009)
 [arXiv:0901.3775 [hep-th]].

\bibitem{general}
J.~Kluson,
  ``Branes at Quantum Criticality,''
 JHEP {\bf 0907}, 079 (2009)
  [arXiv:0904.1343 [hep-th]];\\
H.~Nastase,
  ``On IR solutions in Horava gravity theories,''
  arXiv:0904.3604 [hep-th];\\
D.~Orlando and S.~Reffert,
  ``On the Renormalizability of Horava-Lifshitz-type Gravities,''
  Class.\ Quant.\ Grav.\  {\bf 26}, 155021 (2009)
  [arXiv:0905.0301 [hep-th]];\\
  F.~W.~Shu and Y.~S.~Wu,
  ``Stochastic Quantization of the Ho\v{r}ava Gravity,''
  arXiv:0906.1645 [hep-th];\\
  A.~Kobakhidze,
  ``On the infrared limit of Horava's gravity with the global Hamiltonian
  constraint,''
  arXiv:0906.5401 [hep-th];\\
T.~Harko, Z.~Kovacs and F.~S.~N.~Lobo,
  ``Solar system tests of Ho\v{r}ava-Lifshitz gravity,''
  arXiv:0908.2874 [gr-qc].

\bibitem{cosmology}
T.~Takahashi and J.~Soda,
  ``Chiral Primordial Gravitational Waves from a Lifshitz Point,''
 Phys.\ Rev.\ Lett.\  {\bf 102}, 231301 (2009)
  [arXiv:0904.0554 [hep-th]];\\
G.~Calcagni,
  ``Cosmology of the Lifshitz universe,''
  JHEP {\bf 0909}, 112 (2009)
  [arXiv:0904.0829 [hep-th]];\\
E.~Kiritsis and G.~Kofinas,
  ``Ho\v{r}ava-Lifshitz Cosmology,''
 Nucl.\ Phys.\  B {\bf 821}, 467 (2009)
  [arXiv:0904.1334 [hep-th]];\\
S.~Mukohyama,
  ``Dark matter as integration constant in Horava-Lifshitz gravity,''
 Phys.\ Rev.\  D {\bf 80}, 064005 (2009)
  [arXiv:0905.3563 [hep-th]];\\
S.~Mukohyama,
  ``Caustic avoidance in Horava-Lifshitz gravity,''
  JCAP {\bf 0909}, 005 (2009)
  [arXiv:0906.5069 [hep-th]];\\
  N.~Afshordi,
  ``Cuscuton and low energy limit of Horava-Lifshitz gravity,''
  Phys.\ Rev.\  D {\bf 80}, 081502 (2009)
  [arXiv:0907.5201 [hep-th]];\\
  S.~Carloni, E.~Elizalde and P.~J.~Silva,
  ``An analysis of the phase space of Horava-Lifshitz cosmologies,''
  arXiv:0909.2219 [hep-th];\\
 G.~Leon and E.~N.~Saridakis,
  ``Phase-space analysis of Horava-Lifshitz cosmology,''
  JCAP {\bf 0911}, 006 (2009)
  [arXiv:0909.3571 [hep-th]].

\bibitem{Saridakis2}
C.~Bogdanos and E.~N.~Saridakis,
  ``Perturbative instabilities in Horava gravity,''
  arXiv:0907.1636 [hep-th].

\bibitem{Charmousis}
C.~Charmousis, G.~Niz, A.~Padilla and P.~M.~Saffin,
  ``Strong coupling in Horava gravity,''
  JHEP {\bf 0908}, 070 (2009)
  [arXiv:0905.2579 [hep-th]].

\bibitem{Blas}
D.~Blas, O.~Pujolas and S.~Sibiryakov,
  ``On the Extra Mode and Inconsistency of Horava Gravity,''
  JHEP {\bf 0910}, 029 (2009)
  [arXiv:0906.3046 [hep-th]];\\
K.~Koyama and F.~Arroja,
  ``Pathological behaviour of the scalar graviton in Ho\v{r}ava-Lifshitz
  gravity,''
  arXiv:0910.1998 [hep-th].

\bibitem{us}
X.~Gao, Y.~Wang, R.~Brandenberger and A.~Riotto,
  ``Cosmological Perturbations in Ho\v{r}ava-Lifshitz Gravity,''
  arXiv:0905.3821 [hep-th].

\bibitem{also}
 S.~Mukohyama,
  ``Scale-invariant cosmological perturbations from Ho\v{r}ava-Lifshitz gravity
  without inflation,''
  JCAP {\bf 0906}, 001 (2009)
  [arXiv:0904.2190 [hep-th]];\\
Y.~S.~Piao,
   ``Primordial Perturbation in Horava-Lifshitz Cosmology,''
   Phys.\ Lett.\  B {\bf 681}, 1 (2009)
  [arXiv:0904.4117 [hep-th]];\\
X.~Gao,
  ``Cosmological Perturbations and Non-Gaussianities in Ho\v{r}ava-Lifshitz
  Gravity,''
  arXiv:0904.4187 [hep-th];\\
R.~G.~Cai, B.~Hu and H.~B.~Zhang,
  ``Dynamical Scalar Degree of Freedom in Horava-Lifshitz Gravity,''
 Phys.\ Rev.\  D {\bf 80}, 041501 (2009)
  [arXiv:0905.0255 [hep-th]];\\
B.~Chen, S.~Pi and J.~Z.~Tang,
  ``Scale Invariant Power Spectrum in Ho\v{r}ava-Lifshitz Cosmology without
  Matter,''
  arXiv:0905.2300 [hep-th];\\
Y.~W.~Kim, H.~W.~Lee and Y.~S.~Myung,
  ``Nonpropagation of scalar in the deformed Ho\v{r}ava-Lifshitz gravity,''
  arXiv:0905.3423 [hep-th];\\
  K.~Yamamoto, T.~Kobayashi and G.~Nakamura,
  ``Breaking the scale invariance of the primordial power spectrum in
  Horava-Lifshitz Cosmology,''
  Phys.\ Rev.\  D {\bf 80}, 063514 (2009)
  [arXiv:0907.1549 [astro-ph.CO]];\\
  T.~Kobayashi, Y.~Urakawa and M.~Yamaguchi,
  ``Large scale evolution of the curvature perturbation in Horava-Lifshitz
  cosmology,''
  arXiv:0908.1005 [astro-ph.CO];\\
  P.~Wu and H.~W.~Yu,
  ``Stability of the Einstein static universe in Ho\v{r}ava-Lifshitz gravity,''
  arXiv:0909.2821 [gr-qc];\\
M.~i.~Park,
  ``Remarks on the Scalar Graviton Decoupling and Consistency of Ho\v{r}ava
  Gravity,''
  arXiv:0910.1917 [hep-th].

\bibitem{Maartens}
A.~Wang and R.~Maartens,
  ``Linear perturbations of cosmological models in the Horava-Lifshitz theory
  of gravity without detailed balance,''
  arXiv:0907.1748 [hep-th];\\
  A.~Wang, D.~Wands and R.~Maartens,
  ``Scalar field perturbations in Horava-Lifshitz cosmology,''
  arXiv:0909.5167 [hep-th].

\bibitem{Blas2}
D.~Blas, O.~Pujolas and S.~Sibiryakov,
  ``A healthy extension of Horava gravity,''
  arXiv:0909.3525 [hep-th];\\
B.~Chen, S.~Pi and J.~Z.~Tang,
  ``Power spectra of scalar and tensor modes in modified Horava-Lifshitz
  gravity,''
  arXiv:0910.0338 [hep-th].

\bibitem{RHB}
R.~Brandenberger,
   ``Matter Bounce in Horava-Lifshitz Cosmology,''
Phys.\ Rev.\  D {\bf 80}, 043516 (2009)
  [arXiv:0904.2835 [hep-th]].

\bibitem{Cai:2009in}
  Y.~F.~Cai and E.~N.~Saridakis,
  ``Non-singular cosmology in a model of non-relativistic gravity,''
  JCAP {\bf 0910}, 020 (2009)
  [arXiv:0906.1789 [hep-th]].


\bibitem{Wands1}
D.~Wands,
  ``Duality invariance of cosmological perturbation spectra,''
  Phys.\ Rev.\  D {\bf 60}, 023507 (1999)
  [arXiv:gr-qc/9809062].

\bibitem{Fabio}
F.~Finelli and R.~Brandenberger,
  ``On the generation of a scale-invariant spectrum of adiabatic  fluctuations
  in cosmological models with a contracting phase,''
  Phys.\ Rev.\  D {\bf 65}, 103522 (2002)
  [arXiv:hep-th/0112249].

\bibitem{Wands2}
 L.~E.~Allen and D.~Wands,
  ``Cosmological perturbations through a simple bounce,''
  Phys.\ Rev.\  D {\bf 70}, 063515 (2004)
  [arXiv:astro-ph/0404441].

\bibitem{recent}
Y.~F.~Cai, T.~t.~Qiu, R.~Brandenberger and X.~m.~Zhang,
  ``A Nonsingular Cosmology with a Scale-Invariant Spectrum of Cosmological
  Perturbations from Lee-Wick Theory,''
  Phys.\ Rev.\  D {\bf 80}, 023511 (2009)
  [arXiv:0810.4677 [hep-th]];\\
Y.~F.~Cai, T.~Qiu, R.~Brandenberger, Y.~S.~Piao and X.~Zhang,
  ``On Perturbations of Quintom Bounce,''
  JCAP {\bf 0803}, 013 (2008)
  [arXiv:0711.2187 [hep-th]].

\bibitem{RHBrev}
R.~H.~Brandenberger,
  ``Alternatives to Cosmological Inflation,''
  arXiv:0902.4731 [hep-th].

\bibitem{bounceNG}
Y.~F.~Cai, W.~Xue, R.~Brandenberger and X.~Zhang,
  ``Non-Gaussianity in a Matter Bounce,''
  JCAP {\bf 0905}, 011 (2009)
  [arXiv:0903.0631 [astro-ph.CO]].

\bibitem{Miao}
M.~Li and Y.~Pang,
  ``A Trouble with Ho\v{r}ava-Lifshitz Gravity,''
  JHEP {\bf 0908}, 015 (2009)
  [arXiv:0905.2751 [hep-th]].

\bibitem{Silke}
T.~Sotiriou, M.~Visser and S.~Weinfurtner,
  ``Phenomenologically viable Lorentz-violating quantum gravity,''
 Phys.\ Rev.\ Lett.\  {\bf 102}, 251601 (2009)
  [arXiv:0904.4464 [hep-th]];\\
T.~P.~Sotiriou, M.~Visser and S.~Weinfurtner,
  ``Quantum gravity without Lorentz invariance,''
 JHEP {\bf 0910}, 033 (2009)
  [arXiv:0905.2798 [hep-th]].

\bibitem{MFB}
V.~F.~Mukhanov, H.~A.~Feldman and R.~H.~Brandenberger,
  ``Theory of cosmological perturbations. Part 1. Classical perturbations. Part
  2. Quantum theory of perturbations. Part 3. Extensions,''
  Phys.\ Rept.\  {\bf 215}, 203 (1992).

\bibitem{RHBrev1}
R.~H.~Brandenberger,
  ``Lectures on the theory of cosmological perturbations,''
  Lect.\ Notes Phys.\  {\bf 646}, 127 (2004)
  [arXiv:hep-th/0306071].

\bibitem{SasMuk}
M.~Sasaki,
  ``Large Scale Quantum Fluctuations in the Inflationary Universe,''
  Prog.\ Theor.\ Phys.\  {\bf 76}, 1036 (1986);\\
V.~F.~Mukhanov,
  ``Quantum Theory of Gauge Invariant Cosmological Perturbations,''
  Sov.\ Phys.\ JETP {\bf 67}, 1297 (1988)
  [Zh.\ Eksp.\ Teor.\ Fiz.\  {\bf 94N7}, 1 (1988)].

\bibitem{FF}
F.~Finelli and R.~H.~Brandenberger,
  ``Parametric amplification of gravitational fluctuations during  reheating,''
  Phys.\ Rev.\ Lett.\  {\bf 82}, 1362 (1999)
  [arXiv:hep-ph/9809490].

\bibitem{Zhang}
W.~B.~Lin, X.~H.~Meng and X.~M.~Zhang,
  ``Adiabatic gravitational perturbation during reheating,''
  Phys.\ Rev.\  D {\bf 61}, 121301 (2000)
  [arXiv:hep-ph/9912510].

\bibitem{HV}
  J.~C.~Hwang and E.~T.~Vishniac,
  ``Gauge-invariant joining conditions for cosmological perturbations,''
  Astrophys.\ J.\  {\bf 382}, 363 (1991).

\bibitem{DM}
 N.~Deruelle and V.~F.~Mukhanov,
  ``On matching conditions for cosmological perturbations,''
  Phys.\ Rev.\  D {\bf 52}, 5549 (1995)
  [arXiv:gr-qc/9503050].

\bibitem{Durrer}
R.~Durrer and F.~Vernizzi,
  ``Adiabatic perturbations in pre big bang models: Matching conditions and
  scale invariance,''
  Phys.\ Rev.\  D {\bf 66}, 083503 (2002)
  [arXiv:hep-ph/0203275].

\bibitem{done}
D.~H.~Lyth,
  ``The primordial curvature perturbation in the ekpyrotic universe,''
  Phys.\ Lett.\  B {\bf 524}, 1 (2002)
  [arXiv:hep-ph/0106153];\\
R.~Brandenberger and F.~Finelli,
  ``On the spectrum of fluctuations in an effective field theory of the
  ekpyrotic universe,''
  JHEP {\bf 0111}, 056 (2001)
  [arXiv:hep-th/0109004];\\
  D.~H.~Lyth,
  ``The failure of cosmological perturbation theory in the new ekpyrotic
  scenario,''
  Phys.\ Lett.\  B {\bf 526}, 173 (2002)
  [arXiv:hep-ph/0110007];\\
  J.~c.~Hwang,
  ``Cosmological structure problem in the ekpyrotic scenario,''
  Phys.\ Rev.\  D {\bf 65}, 063514 (2002)
  [arXiv:astro-ph/0109045].

\bibitem{Ekp}
J.~Khoury, B.~A.~Ovrut, P.~J.~Steinhardt and N.~Turok,
  ``The ekpyrotic universe: Colliding branes and the origin of the hot big
  bang,''
  Phys.\ Rev.\  D {\bf 64}, 123522 (2001)
  [arXiv:hep-th/0103239].

\bibitem{BFS}
 R.~Brandenberger, H.~Firouzjahi and O.~Saremi,
  ``Cosmological Perturbations on a Bouncing Brane,''
  JCAP {\bf 0711}, 028 (2007)
  [arXiv:0707.4181 [hep-th]].

\bibitem{ABB}
 S.~Alexander, T.~Biswas and R.~H.~Brandenberger,
  ``On the Transfer of Adiabatic Fluctuations through a Nonsingular
  Cosmological Bounce,''
  arXiv:0707.4679 [hep-th].

\bibitem{RHBcyclic}
R.~H.~Brandenberger,
  ``Processing of Cosmological Perturbations in a Cyclic Cosmology,''
  Phys.\ Rev.\  D {\bf 80}, 023535 (2009)
  [arXiv:0905.1514 [hep-th]].

\end{thebibliography}
\end{document}